\documentclass[sigconf]{acmart}

\copyrightyear{2026}
\acmYear{2026}
\setcopyright{cc}
\setcctype{by}
\acmConference[KDD '26]{Proceedings of the 32nd ACM SIGKDD Conference on Knowledge Discovery and Data Mining V.2}{August 09--13, 2026}{Jeju Island, Republic of Korea}
\acmBooktitle{Proceedings of the 32nd ACM SIGKDD Conference on Knowledge Discovery and Data Mining V.2 (KDD '26), August 09--13, 2026, Jeju Island, Republic of Korea}
\acmDOI{10.1145/3770855.3818320}
\acmISBN{979-8-4007-2259-2/2026/08}

\usepackage{multirow}
\usepackage[utf8]{inputenc}
\usepackage{xcolor}
\usepackage{colortbl}
\usepackage{enumitem}

\begin{document}

\title{Atomic Intent Reasoning: Bringing LLM Semantics to Industrial Cross-Domain Recommendations}

\author{Zhuohang Jiang}
\authornote{Work done during an internship at Kuaishou Technology.}
\authornote{Authors contributed equally to this research.}
\affiliation{%
  \institution{The Hong Kong Polytechnic University}
  \city{Hong Kong SAR}
  \country{China}
}
\email{zhuohang.jiang@connect.polyu.hk}

\author{Yuxin Chen}
\authornotemark[2]
\affiliation{%
  \institution{Kuaishou Technology}
  \city{Beijing}
  \country{China}
}
\email{chenyuxin06@kuaishou.com}

\author{Shijie Wang}
\affiliation{%
  \institution{The Hong Kong Polytechnic University}
  \city{Hong Kong SAR}
  \country{China}
}
\email{shijie.wang@connect.polyu.hk}

\author{Haohao Qu}
\affiliation{%
  \institution{The Hong Kong Polytechnic University}
  \city{Hong Kong SAR}
  \country{China}
}
\email{haohao.qu@connect.polyu.hk}

\author{Jindong Zhou}
\affiliation{%
  \institution{Kuaishou Technology}
  \city{Beijing}
  \country{China}
}
\email{zhoujindong@kuaishou.com}

\author{Wenqi Fan}
\authornote{Corresponding Author.}
\affiliation{%
  \institution{The Hong Kong Polytechnic University}
  \city{Hong Kong SAR}
  \country{China}
}
\email{wenqifan03@gmail.com}

\author{Qing Li}
\authornotemark[3]
\affiliation{%
  \institution{The Hong Kong Polytechnic University}
  \city{Hong Kong SAR}
  \country{China}
}
\email{csqli@comp.polyu.edu.hk}

\author{Dongxu Liang}
\affiliation{%
  \institution{Kuaishou Technology}
  \city{Beijing}
  \country{China}
}
\email{liangdongxu@kuaishou.com}

\author{Jun Wang}
\authornotemark[3]
\affiliation{%
  \institution{Kuaishou Technology}
  \city{Beijing}
  \country{China}
}
\email{wangjun03@kuaishou.com}

\renewcommand{\shortauthors}{Zhuohang Jiang et al.}

\begin{abstract}

Cross-domain recommendation is an essential problem in content-to-merchant platforms. Its objective is to leverage user interactions with content to infer potential purchasing intent on the merchant side, thereby enhancing conversion rates and commercial value. However, in real industrial scenarios, cross-domain recommendation faces multiple challenges: significant semantic gaps exist between different domains, and user cross-domain behavior sequences are often massive in scale and rich in noise. 
Although large language models (LLMs) offer powerful semantic understanding and reasoning capabilities to alleviate semantic gap issues, their inference costs remain prohibitive under critical online inference latency constraints.
To address these issues, this paper introduces \ourname{} (Atomic Intent Reasoning), an LLM-driven cross-domain recommendation framework designed for industrial-grade deployment. By migrating LLM inference to the offline phase and dynamically constructing user intent representations via efficient retrieval and composition during online operations, it achieves a \textbf{400×} throughput gain over real-time LLM invocation while preserving semantic consistency.
Also performing structured modeling and goal-aware compression on lengthy, heterogeneous cross-domain behavior sequences, transforming into compact evidence representations highly relevant to candidate products for CTR prediction and refined ranking. Experimental results across multiple public datasets demonstrate that our method achieves state-of-the-art performance in cross-domain recommendation tasks. Furthermore, large-scale online A/B testing conducted in Kuaishou E-commerce's real-world business scenarios shows that our approach delivers stable and significant improvements across multiple core business metrics, including a \textbf{+3.446\%} increase in GMV, fully validating its effectiveness and practical value in industrial-scale recommendation systems.

\end{abstract}



\newcommand{\ourname}{\textbf{\textsc{AIR}}}

\maketitle
\section{Introduction}

\begin{figure}
    \centering
    \includegraphics[width=1.0\linewidth]{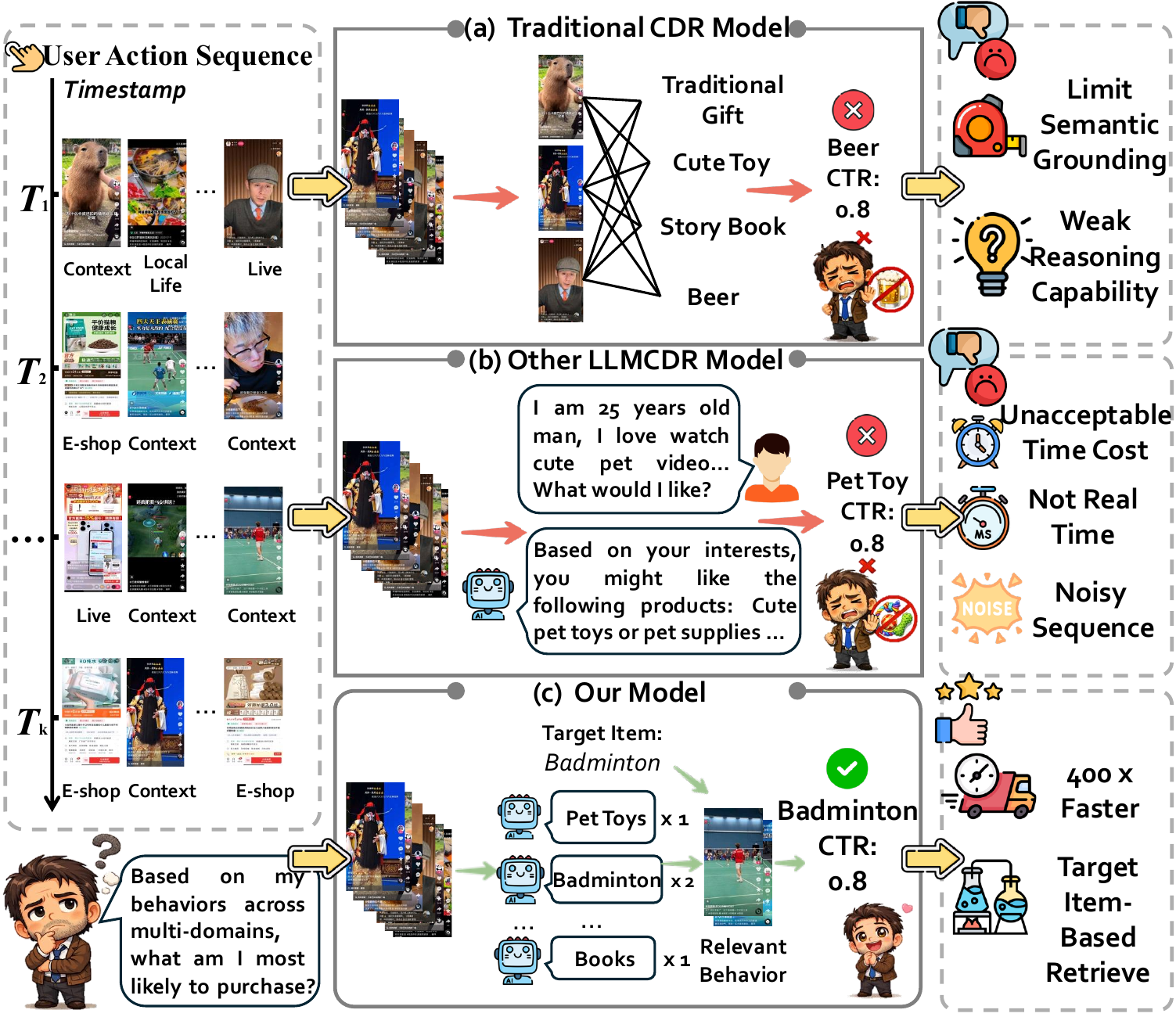}
    \caption{Kuaishou spans multiple heterogeneous domains, making cross-domain recommendation crucial for intent transfer, but facing two key challenges:
    (1) Prohibitive Real-Time Inference Expanse of Large Language Models, and (2) Large-Scale and Noisy User Behavior Sequences.}
    \label{fig:CDSR}
    \vskip -0.1in
\end{figure}

Driven by the rapid proliferation of internet services and mobile applications, recommender systems~\cite{hu2026stop, pan2026beyond, zhang2024linear,fan2018deep,fan2022graph} have become a pivotal component in alleviating information overload and influencing users' decision-making (e.g., clicks, add-to-cart actions,  purchases) across various online platforms, such as Kuaishou, Taobao, and Xiaohongshu. 
For example, on platforms like Kuaishou and TikTok, users' online behaviors increasingly span multiple scenarios, such as watching live streams on the content side and browsing or purchasing products on the e-commerce side, thereby forming a tightly coupled content-to-commerce loop.
This coupling motivates cross-domain recommendation (CDR), which leverages users’ content-side online behaviors to capture their personalized preferences for items (i.e., products) and predict users' intent for the next item, thereby increasing Gross Merchandise Value (GMV) on the e-commerce side.
As Figure~\ref{fig:CDSR}~(a) shows, most existing cross-domain recommendation methods rely on ID-level collaborative transfer~\cite{zang2022survey} (e.g., shared embeddings~\cite{zhao2023cross, cao2023towards} or GNNs~\cite{xie2022contrastive,wang2025graph,fan2019graph,fan2020graph}) to connect entities across scenarios, facilitating the learning of users' personalized preferences for the next item (i.e., user intent modeling). 
While effective at capturing co-occurrence correlations to understand users' preferences, these approaches provide limited semantic grounding and thus often struggle to capture fine-grained intent transitions and accurately infer purchase-oriented preferences from content-side behaviors.

Large language models (LLMs)~\cite{patil2024review, minaee2024large-023,ni2026streasoner,xu2026comprehensive} have recently demonstrated strong semantic understanding and contextual reasoning~\cite{jiang2025hibench}, making them a natural candidate to bridge the semantic shift between content engagement and commerce conversion.
Consequently, an increasing number of studies have incorporated LLMs to inject semantic signals into user modeling, such as interpreting behaviors in natural language, transferring preference cues across domains, or leveraging LLM-derived representations as auxiliary supervision~\cite{petruzzelli2024instructing, xin2025llmcdsr, 10.1145/3746027.3755347}, as illustrated in Figure~\ref{fig:CDSR}~(b).
However, their effectiveness in real-world content-to-commerce systems remains limited. 
Performance gains on academic benchmarks do not reliably transfer to industrial traffic with highly heterogeneous, rapidly evolving behaviors.
Additionally, industrial systems typically have massive multi-scenario logs, where the core bottleneck is filtering intent-bearing behaviors from extremely long and noisy histories.

Satisfying these requirements in real-world content-to-commerce systems presents significant challenges.
\textbf{First}, online LLM inference is prohibitively expensive under critical online inferece latency constrains (milliseconds), and periodic offline updates of user profiles fail to capture rapidly evolving user interests, compromising recommendation timeliness.
\textbf{Second}, user behavior histories are typically extensive, heterogeneous, and noisy; among numerous cross-scenario events, merely a limited subset provides direct evidence for a given target item.
Feeding raw sequences into the model not only amplifies semantic noise but also incurs substantial computation, making efficient target-aware behavior selection essential for scalable and accurate prediction.

To address these industrial challenges, we propose \ourname{} (\textbf{A}tomic \textbf{I}ntent \textbf{R}easoning), a cross-domain recommendation framework for short-video and e-commerce scenarios, facilitating LLM-level semantic reasoning under critical latency online serving constraints. As illustrated in Figure~\ref{fig:CDSR}~(c), the proposed \ourname{} follows an offline to online pipeline. 
In the offline stage, \ourname{} leverages LLMs to transform user events, together with user attributes and object descriptions, into atomic behavior intent units, which are subsequently organized in an intent knowledge base for high-throughput retrieval. During online inference stage, the system retrieves and composes these cached atomic intents from the user’s recent behaviors to construct an up-to-date intent representation without invoking LLMs, thereby substantially reducing serving latency while retaining semantic signals. To further address the challenges posed by lengthy and noisy multi-scenario behavior histories, \ourname{} constructs a unified user intent tree and conducts target-aware semantic retrieval with respect to the current target item, yielding a compact set of highly relevant and evidential intents for prediction. Finally, a multi-head attention module fuses the retrieved target-related intents with complementary user-interest cues, enabling fine-grained preference modeling for more accurate and robust personalized recommendations.

In this paper, our proposed \ourname{} maintain the semantic understanding advantages of large language models while achieving the high efficiency, high concurrency, and low latency required for industrial deployment through structural reconstruction and process decoupling. 
Specifically, when validated in Kuaishou E-commerce’s real-world operations \footnote{https://www.kuaishou.com/}, our approach significantly outperforms traditional cross-domain recommendation methods across multiple core metrics, including Paid Order Count, GPM, OPM, and GMV, validating its practical value in large-scale industrial systems.

The main contributions of this work are summarized as follows: 
\begin{itemize} 

    \item 
    We propose \textbf{\ourname{}} (Atomic Intent Reasoning), an LLM-powered cross-domain recommendation framework for content-to-commerce systems, which achieves effectiveness comparable to real-time LLM calls while maintaining millisecond-level serving latency, and delivers SOTA performance on public benchmarks.
    
    \item 
    We design an \textbf{atomic intent caching and composition} mechanism: user events are distilled into atomic behavior--intent units offline and served via a high-throughput intent knowledge base, yielding an approximately \textbf{400$\times$ throughput gain} over real-time LLM invocation.
    
    \item
    We propose \textbf{target-aware intent retrieval} to extract compact, high-evidence intents from long and noisy multi-scenario histories, and use \textbf{multi-head attention} to fuse target-related intents with other user-interest signals for fine-grained preference modeling. 
    
    \item Extensive experiments on public benchmarks show the effectiveness of the proposed framework, 
    and an industrial A/B test further demonstrates consistent online gains of \textbf{+3.446\%} on \textit{GMV}.
\end{itemize}

\section{Related Works}

\noindent\textbf{LLM-based Recommendation.}
Large language models (LLMs)~\cite{jiang2025hibench} scaled to billions of parameters demonstrate exceptional capabilities in language understanding, generation, and reasoning, exhibiting strong generalization to downstream tasks and domains \cite{yuan2025mkg, ning2025survey, deng2026fundus, team2026venus}. Leveraging these advancements, LLM-empowered recommender systems have attracted significant research interest, offering novel opportunities to advance the field \cite{wang2025knowledge}.
To effectively adapt LLMs for recommendation, current research primarily explores several paradigms—such as pre-training, fine-tuning, prompting, retrieval, and reinforcement learning~\cite{fan2025towards, jiang2025qa, jiang2026superglasses}. These paradigms aim to bridge the semantic understanding capabilities of LLMs with the personalized preference modeling requirements of recommender systems. Representative approaches include P5~\cite{geng2022recommendation} and TokenRec~\cite{qu2025tokenrec,qu2025generative}, which employ prompt-based learning to unify diverse recommendation tasks under a language modeling framework and achieve notable zero-shot generalization through personalized prompting~\cite{jiang2026superglasses}. 
However, despite their promise, these pioneering methods remain difficult to deploy at an industrial scale, especially under critical online inference latency constraints, owing to their inadequate computational efficiency~\cite{wang2025rethinking} and limited ability to model cross-domain user intent~\cite{zang2022survey}.

\noindent\textbf{Cross-Domain Recommendation.}
Cross-domain Recommendation (CDR) models~\cite{chen2024survey} capture user preferences by leveraging users’ interaction behaviors across multiple domains.
Traditional recommendation methods~\cite{guo2021gcn,cao2022contrastive}, typically designed for single-domain behaviors, struggle with data sparsity and fail to capture interest transfer across different and heterogeneous scenarios. 
Existing CDR approaches incorporate cross-domain information through shared representations, alignment mechanisms~\cite{cao2022contrastive}, and graph-based modeling~\cite{guo2021gcn}. However, these methods primarily capture shallow co-occurrence patterns and lack the capacity to model the underlying semantic and logical dependencies among user behaviors.
Recently, Large Language Models (LLMs)~\cite{bai2023qwen-1e4,minaee2024large-023} have been increasingly explored for cross-domain recommendation, due to their remarkable abilities in semantic representation and reasoning. Representative methods, such as LLM4CDSR~\cite{xin2025llmcdsr} and LLMCDSR~\cite{liu2025bridge}, leverage LLMs to unify item representations or generate cross-domain pseudo-interactions, thereby alleviating overlap and knowledge-transfer issues.
However, existing LLM-based approaches primarily operate at a coarse-grained semantic level, which limits their ability to capture fine-grained sequential dependencies and dynamic interest evolution, especially in industrial settings characterized by large-scale, heterogeneous, and highly complex user behavior data.

\section{Method}

\begin{figure*}[t]
    \centering
    \includegraphics[width=1\linewidth]{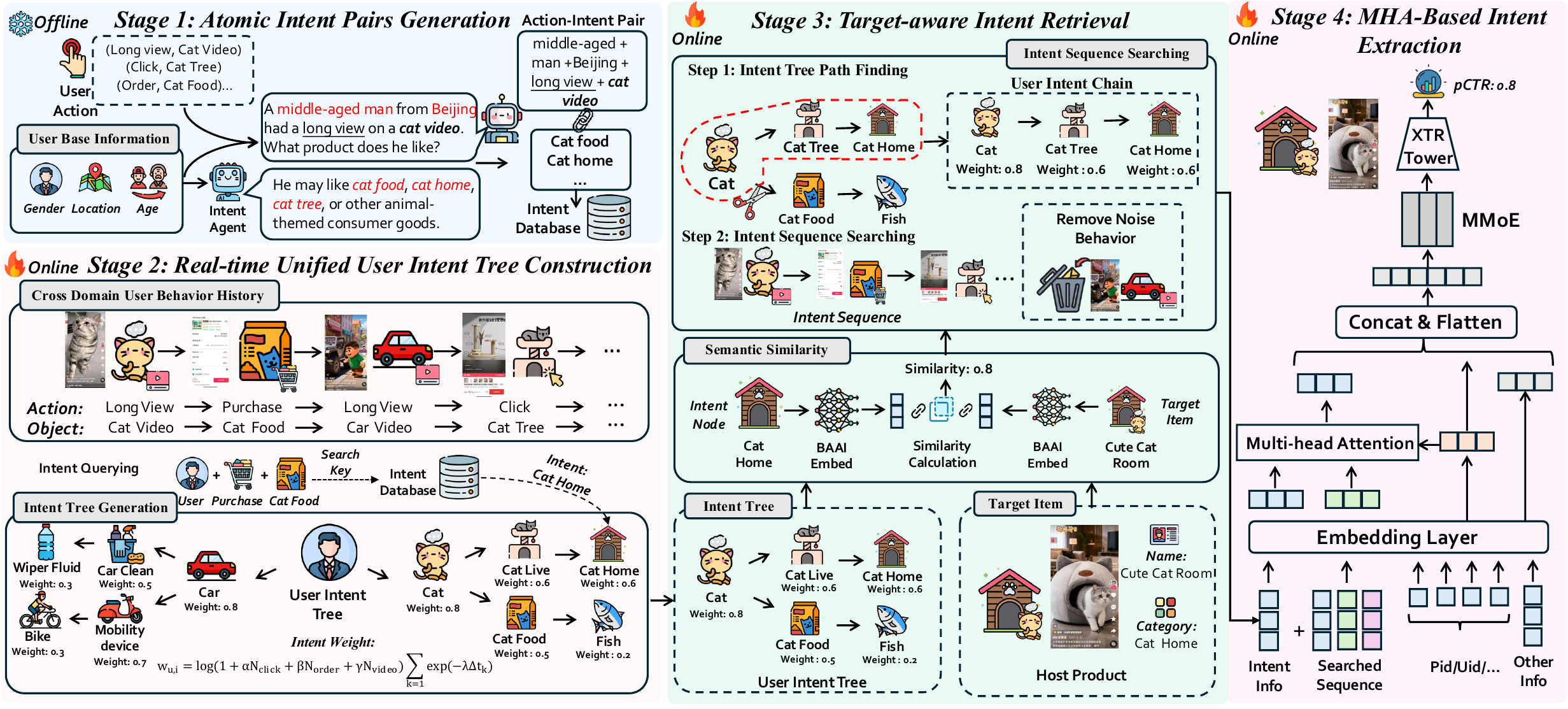}
    \vskip -0.1in
    \caption{Our proposed \ourname{} (Atomic Intent Reasoning) tackles cross-domain recommendation via a four-stage pipeline: (1) Atomic Intent Pair Generation, (2) Real-time Unified User Intent Tree Construction, (3) Target-aware Intent Retrieval, and (4) MHA-based Intent Extraction.}
    \label{fig:model}
\end{figure*}

\subsection{Overall Architecture}

To operationalize LLM-derived semantic understanding under industrial latency constraints, we design AIR as a decoupled offline-to-online framework that transforms raw cross-domain behaviors into target-relevant intent representations.
As illustrated in Figure~\ref{fig:model}, the proposed \ourname{} includes four stages: 1) \textbf{Atomic Intent Pair Generation} for offline semantic grounding; 2) \textbf{Real-time Unified User Intent Tree Construction} for behavior sequence structuring; 3) \textbf{Target-aware Intent Retrieval} for relevance refinement; and 4) \textbf{MHA-based Intent Extraction} for holistic preference fusion, which together enable fine-grained user preference modeling.
Specifically, the offline stage leverages LLMs to construct atomic behavior-intent pairs that encode fine-grained user semantics and stores them in a high-concurrency intent knowledge base.
During online serving, these atomic behavior-intent pairs are efficiently retrieved and dynamically assembled into a real-time unified user intent tree, transforming massive, heterogeneous behavior sequences into structured semantic representations.
Conditioned on the target item, the framework further performs target-aware retrieval over the intent tree to distill compact and high-evidence behavioral subsequences that are most informative for downstream prediction.
A multi-head attention module integrates target-related intent representations with complementary user-interest signals for fine-grained preference modeling.
This four-stage framework fully decouples LLM inference from online serving while preserving real-time semantic awareness, thereby enabling low-latency, high-throughput recommendation in industrial-scale systems.

\subsection{Atomic Intent Pair Generation}
To address the \textit{prohibitive real-time inference expense of LLMs}, we propose an \textbf{offline generation method for atomic user intent pairs}, which decouples LLM reasoning from the online recommendation pipeline. 
Notably, the latent intent underlying complex multi-behavior sequences can be effectively approximated by composing single-behavior atomic intents.
As illustrated in Stage 1 of Figure~\ref{fig:model}, cross-domain user interaction sequences are decomposed into atomic behavioral units, each mapped to a fine-grained intent representation. By composing these atomic intents, the framework constructs a unified semantic representation of the user’s latent preference evolution, which serves as an efficient approximation to holistic sequence-level semantic reasoning. Such a design enables real-time emulation of LLM-driven inference through offline intent caching and lightweight online retrieval.

Mathematically, we define each user event as: $e = (a, o, u)$, where $a$ denotes the action type (e.g., \textit{click}, \textit{long-view}, and \textit{purchase}), $o$ is the behavior object (e.g., video or product), and $u = (\text{gender}, \text{age}, \text{location})$ denotes the user profile attributes.
The user event $(a,o,u)$ is fed into an LLM prompt that combines (i) user context, (ii) behavioral semantics, and (iii) object description.
The LLM outputs a set of hierarchical intent paths:
\begin{equation}
\mathcal{P}(e) = \{ p_1, \dots, p_m \}, \quad \forall p = [c_1 \rightarrow c_2 \rightarrow \dots \rightarrow c_k],
\end{equation}
where $m$ and $k$ denote the potential number of user intents and their category number, respectively.
Each $c_\ell$ represents an intent category at depth $\ell$, forming a coarse-to-fine intent trajectory.

To support scalable real-time inference, we construct action-specific intent caches:
\begin{equation}
\mathcal{P}_a = \bigcup_{e:\,a_e = a} \mathcal{P}(e),
\label{eq:intent_caches}
\end{equation}
allowing heterogeneous action types to contribute with differentiated importance in downstream modeling.
A comprehensive set of atomic intent pairs is thus derived and subsequently utilized for retrieval and intent tree construction, as detailed in the next section.

\subsection{Real-time Unified User Intent Tree Construction}

To more effectively characterize the multi-level structure of user intent and the differentiated importance of heterogeneous behavior types, the cached action-specific intent paths are aggregated into a unified \textbf{action-related Intent Tree}. This hierarchical structure captures both the layered organization of user interests and the varying strengths of cross-behavior signals, yielding a semantically expressive and discriminative representation for downstream target-aware retrieval and ranking. Within the intent tree, each node denotes an intent concept at a particular level of abstraction, while edges encode the corresponding hierarchical parent-child relations.
Notably, each node maintains \textbf{action-specific statistics}, enabling differentiated weighting over heterogeneous behavior types.
For a node $u$ (an intent category), we compute its preference weight $w_u$ by combining multi-action counts with temporal decay:
\begin{align}
w_u &= \log(1+N(u)) \sum\nolimits_{k=1}^{n_u}\exp(-\lambda\,\Delta t_k), \\
N(u) &= \alpha N_{\text{click}}(u)+\beta N_{\text{order}}(u)+\gamma N_{\text{view}}(u),
\end{align}
where $N_{\text{click}}(u)$, $N_{\text{order}}(u)$, and $N_{\text{view}}(u)$ are the node hit counts from click/order/video (or long-view) evidence, respectively, and $\alpha$, $\beta$, and $\gamma$ are the corresponding coefficients that balance their contributions.
$\Delta t_k$ is the time gap between the $k$-th supporting event and the current time, and $\lambda$ controls the decay strength.
The logarithmic term stabilizes heavy-tailed frequency, while the decay term emphasizes recent behaviors (i.e., stronger short-term intent).
In addition to the preference weight $w_u$, each node stores its supporting behavior object IDs per action type:
\begin{equation}
\mathcal{I}_a(u) = \{ o_i \mid e_i = (a, o_i, u_i),\; p \in \mathcal{P}(e_i),\; u \in p \},
\end{equation}
where $\mathcal{I}_a(u)$ denotes the set of behavior object IDs (e.g., videos or items) that contribute to node $u$ under action type $a$. These will serve as explicit evidence for target-aware retrieval in later stages.
This intent tree is constructed online via lightweight aggregation over cached intent paths, supporting efficient real-time serving while preserving rich multi-level user preferences.

During serving, the user-level intent representation is dynamically assembled by merging intents from the user’s recent behavior history $\mathcal{H}_{\text{user}}$ corresponding to Eq.~\eqref{eq:intent_caches}:
\begin{equation}
\mathcal{P}_{\text{user}} = \bigcup_{e \in \mathcal{H}_{\text{user}}} \mathcal{P}(e).
\end{equation}
This design eliminates costly online LLM inference while retaining fine-grained intent signals, thereby supporting personalized recommendations in industrial-scale settings.

\begin{figure*}
    \centering
    \includegraphics[width=\linewidth]{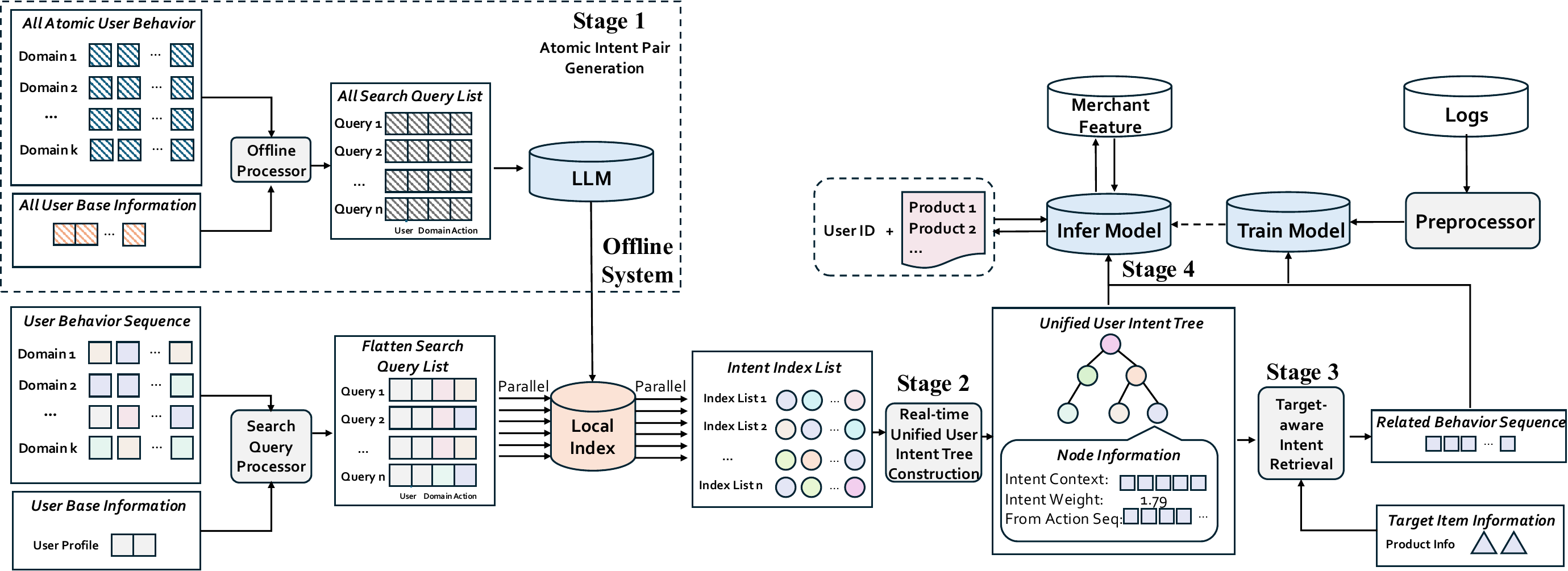}
    \caption{Online serving in \ourname{} with separated offline/online serving stages. Offline, LLMs generate and index atomic behavior–intent pairs. Online, user behaviors are decomposed into parallel intent queries and efficiently retrieved to summarize user intent, achieving an effect equivalent to real-time LLM inference while supporting low-latency, high-QPS recommendation.}
    \label{fig:Implementation}
\end{figure*}

\subsection{Target-aware Intent Retrieval}

Since user behavior sequences typically span multiple domains and exhibit substantial scale and noise, directly encoding the full lifetime history in online models is both computationally prohibitive and susceptible to semantic interference from behaviors irrelevant to the current recommendation target.
Accordingly, \emph{target-aware sequence retrieval} conditioned on a given candidate item $v$ is introduced to derive a compact and precise representation of user intent.
Motivated by this observation, \emph{target-aware reasoning search} is conducted over the intent tree to identify intent representations relevant to the target item and compress the user’s behavioral history into a concise \emph{user intent chain} tailored to $v$.
Specifically, the retrieval framework encompasses the following three critical processes: namely Semantic Representation Construction, Intent Chain Finding, and Intent Sequence Searching.

\noindent\textbf{Semantic Representation Construction.}
Semantic representations are constructed for both the target item and the user intent nodes. Let $\mathbf{e}_v$ denote the embedding of $v$, derived from its title, category, and attributes, optionally incorporating multimodal signals; let $\mathbf{e}_u$ denote the embedding of intent node $u$, derived from its canonical name and semantic expansions. Target-to-intent relevance is measured using semantic similarity:
\begin{equation}
s(u,v)=\cos(\mathbf{e}_u,\mathbf{e}_v).
\end{equation}
Based on this relevance score, a small candidate set of intent nodes is selected from the intent tree $\mathcal{T}$:
\begin{equation}
\mathcal{U}_v=\text{Top-}M_{u\in\mathcal{T}}\, s(u,v).
\end{equation}
For each candidate node $u$, a hierarchical intent chain is constructed by recursively tracing its ancestors to the root, thereby forming a depth-ordered path that preserves the semantic abstraction process from coarse-grained to fine-grained levels.

\noindent\textbf{Intent Chain Finding.}
Candidate intent chains are scored by jointly considering target relevance and user preference strength. A simple yet effective scoring function is defined as
\begin{equation}
\text{score}(u;v)= s(u,v)\cdot g(w_u),
\end{equation}
where $w_u$ denotes the accumulated user preference weight on intent node $u$, and $g(\cdot)$ is a monotonic mapping function such as $g(w)=\log(1+w)$. The top-ranked chains are retained and merged with de-duplication to obtain a short, hierarchical {user intent chain} $\mathcal{C}_v$. This representation exhibits strong interpretability, as each node is semantically aligned with the target item and grounded in preference-aware user behavioral evidence.

\noindent\textbf{Intent Sequence Searching.}
Given the chain skeleton, supporting evidence is subsequently retrieved from cached behavior-specific sources.
Specifically, for each node in the chain, a compact set of supporting behavioral instances, such as clicks, orders, or video interactions, is retrieved, followed where appropriate by lightweight denoising to suppress irrelevant signals. Such denoising may include (i) discarding evidence with low semantic similarity to the target item, (ii) filtering temporally stale evidence through time-decay mechanisms, and (iii) constraining the evidence budget associated with each node.
A compact sequence of intent tokens, together with their associated supporting evidence, is thereby derived to provide highly informative conditioning signals for downstream CTR prediction and fine-grained ranking, while ensuring bounded online computational complexity.

\subsection{MHA-based Intent Extraction}
The retrieved intent chain is further transformed into a target-aware interest representation via \textbf{multi-head attention}.
Let $\mathbf{q}_v$ be the target item embedding, and let $\mathbf{X}=[\mathbf{x}_1;\ldots;\mathbf{x}_L]$ be the embeddings of the $L$ retrieved intent-chain tokens (optionally fused with action/source and time features). Multi-head attention produces a target-conditioned summary:
\begin{equation}
\mathbf{z}_v = \mathrm{MHA}(\mathbf{q}_v, \mathbf{X}, \mathbf{X}).
\end{equation}

This mechanism enables different attention heads to capture complementary facets of user intent, such as coarse-grained category alignment, fine-grained preference patterns, and heterogeneous evidence sources. The resulting representation $\mathbf{z}_v$ is subsequently integrated with standard user and item features and fed into the downstream CTR prediction or ranking model.
Accordingly, LLM-driven semantic transfer is realized through retrieval, while serving-time computation remains lightweight and scalable.

\section{Deployment}

Industrial-scale recommender and advertising systems must serve massive request volumes under strict latency budgets, typically within tens to hundreds of milliseconds, making direct deployment of LLM-based user intent modeling highly challenging. On platforms such as Kuaishou, user behaviors span multiple domains, yielding long and heterogeneous cross-domain sequences that are costly and noisy for direct LLM processing. More importantly, second-level inference latency and the enormous overhead of large-scale offline updates render such approaches impractical for real-time industrial serving.

As depicted in Figure~\ref{fig:Implementation}, \ourname{} facilitates the integration of the LLM’s reasoning capabilities into an industrial online recommender system by decoupling slow semantic inference from real-time serving, thereby enabling equivalent execution within the system.

During offline serving, \ourname{} constructs an industrial-grade LLM-based semantic generation pipeline that extracts latent intent semantics from cross-domain user behaviors. The pipeline processes user profile attributes (e.g., age, gender, region) and diverse historical behaviors through the LLM to infer behavior–intent relationships, which are then atomized into stable, reusable fine-grained intent units for efficient downstream utilization.
These atomic intents are subsequently structured, indexed, and stored in a compact key–value format within local retrieval structures, effectively offloading the computationally intensive LLM inference to the offline stage. This architecture enables efficient retrieval and reduces the real-time processing burden on the online serving pipeline.

During online serving, the Search Query Processor decomposes large, heterogeneous behavior sequences into lightweight intent queries, which are executed in parallel across local intent indices, enabling the retrieval of multiple intent signals with millisecond-level latency.
The retrieved atomic intents are dynamically aggregated into a hierarchical Unified User Intent Tree, which maintains multi-level semantic structures while incorporating action-specific weights and contextual information.
To further align user intent with CTR prediction, the system conducts target-item–aware sequence retrieval, extracting high-evidence behavioral subsequences pertinent to the candidate item, thereby effectively mitigating noise from large-scale, cross-domain user histories.
Consequently, online inference relies solely on lightweight retrieval, parallel aggregation, and tree-based intent composition, bypassing direct LLM execution. This approach ensures the online process is functionally equivalent to real-time LLM reasoning, while fully satisfying the stringent latency, scalability, and high-QPS requirements of industrial-scale recommendation systems.
\section{Experiments}

\subsection{Experimental Settings}
\subsubsection{\textbf{Datasets}}
Comprehensive experiments were conducted on three benchmark datasets to thoroughly evaluate the effectiveness of the proposed \ourname{} framework and its key components.
Following prior work, two CDSR datasets were constructed from the publicly available Amazon dataset\footnote{https://jmcauley.ucsd.edu/data/amazon/index\_2014.html}.
To assess the performance of \ourname{} under different degrees of domain relatedness, four domains were selected to form two cross-domain pairs, namely \emph{Movie--Book} and \emph{Food--Kitchen}.
In addition, the industrial dataset was collected from \emph{Kuaishou}'s e-commerce platform\footnote{https://www.kuaishou.com/}, which serves over 400 million active users worldwide. This dataset comprises merchant short videos and live-stream photos, with both photo-level and item-level e-commerce data spanning tens to hundreds of billions of records.
For the online A/B test, 5.08\% of the platform traffic was allocated to compare the proposed model with the existing production baseline.
Detailed dataset statistics are reported in Table~\ref{tab:Statistics}.

\begin{table}[t]
\vskip -0.1in
\caption{Dataset Statistics.}
\vskip -0.1in
\label{tab:Statistics}
\resizebox{\linewidth}{!}{
\begin{tabular}{cccccc}
\toprule
\textbf{Dataset} & \textbf{Domain} & \textbf{Users} & \textbf{Items} & \textbf{Ave. Seq Length} \\
\midrule
\multirow{2}{*}{Movie-Book} & Movie & 9,485 & 12,875 & \multirow{2}{*}{9.01} \\
& Book & 52,908 & 93,860 & \\
\midrule
\multirow{2}{*}{Food-Kitchen} & Food & 10,822 & 8,661 & \multirow{2}{*}{8.22} \\
& Kitchen & 41,670 & 27,637 & \\
\midrule
\multirow{2}{*}{Industrial} & Merchant & \multirow{2}{*}{0.4 billion} & billions & \multirow{2}{*}{thousands} \\
& Content & & tens of billions & \\
\bottomrule
\end{tabular}
}
\vskip -0.1in
\end{table}

\subsubsection{\textbf{Competitors}}
To assess the performance of our proposed model, we conduct comparisons with two categories of baseline models: (i) six single-domain sequential recommendation (SR) models and (ii) five cross-domain sequential recommendation (CDSR) models.

The baseline single-domain sequential recommendation models considered in this study include
\textbf{FPMC}~\cite{rendle2010factorizing}, which factorizes personalized Markov chains to model sequential transition patterns;
\textbf{Caser}~\cite{tang2018personalized}, which transforms item sequences into image-like representations and applies convolutional filters to capture sequential dependencies;
\textbf{GRU4Rec}~\cite{hidasi2015session}, which employs recurrent neural networks to model session-based sequential interactions;
\textbf{SRGNN}~\cite{wu2019session}, which utilizes graph neural networks to capture item transition structures in session sequences and adopts an attention mechanism to integrate global preference with current interest;
\textbf{FEARec}~\cite{du2023frequency}, which models both low- and high-frequency information through a ramp structure and an attention mechanism for periodic pattern learning; and
\textbf{SASRec}~\cite{kang2018self}, a self-attention-based model that captures long-range sequential dependencies for next-item prediction.

The compared cross-domain sequential recommendation models include
\textbf{TPUF}~\cite{ding2023tpuf}, which transfers pre-trained user features from the source domain via a feature mapping and aggregation framework;
\textbf{$\pi$-Net}~\cite{ma2019pi}, which employs gating mechanisms to filter and transfer information across domains with overlapping users;
\textbf{C2DSR}~\cite{cao2022contrastive}, which combines graph neural networks and self-attention to model intra- and inter-sequence dependencies, while adopting contrastive infomax for cross-domain representation learning;
\textbf{MGCL}~\cite{xu2025multi}, which applies contrastive learning to capture dynamic sequential patterns and complementary preferences across domains; and
\textbf{LLMCDSR}~\cite{xin2025llmcdsr}, which leverages LLM-generated and filtered pseudo cross-domain interactions, together with collaborative signals and meta-learning, to alleviate sparse user overlap and enhance cross-domain sequential recommendation performance.

\begin{table*}[t]
\vskip -0.1in
\caption{ The Overall Recommendation Performance on the Amazon Dataset.}
\vskip -0.1in
\label{tab:Main}
\begin{tabular}{c|cc|cc|cc|cc}
\toprule
\multirow{2}{*}{Method} & \multicolumn{2}{c|}{Movie} & \multicolumn{2}{c|}{Book} & \multicolumn{2}{c|}{Food} & \multicolumn{2}{c}{Kitchen} \\
                        & NDCG@10      & HR@10      & NDCG@10      & HR@10     & NDCG@10      & HR@10     & NDCG@10       & HR@10       \\ \midrule
TPUF                   & 0.035             &       0.068     &    0.029          &     0.059      &        0.017      &    0.040       &      0.012         &      0.025      \\
$\pi$-Net               &    0.039          &     0.077       &      0.044        &       0.089    &     0.032         &      0.063     &        0.036       &      0.078       \\
C2DSR                   &    0.035          &   0.064         &      0.014        &    0.031       &        0.069      &   0.131        &       0.023        &   0.046          \\
MGCL                    &   0.047           &    0.097        &        0.036      &      0.071     &      0.041        &     0.085      &    0.041           &       0.082      \\ 
LLMCDSR                  &     \underline{0.055}         &     \underline{0.107}       &      \underline{0.117}        &    \underline{0.214}       &   \underline{0.221}           &   \underline{0.433}       &         \underline{0.252}     &    0.439   \\
\midrule
FPMC                    &  0.031            &  0.062          &      0.061        &   0.108        &        0.126      &        0.229   &         0.086      &       0.153      \\
Caser                       &    0.031          &  0.057         &  0.053             &  0.102    &       0.058       &      0.133      &      0.085        &     0.166         \\
FEARec                  &        0.030      &    0.062        &       0.063       &    0.109      &       0.052       &     0.107      &  0.068             &    0.125      \\
GRU4Rec                 &          0.034    &      0.070      &       0.068       &   0.126        &    0.118          &    0.232      &        0.130       &      0.247       \\
SRGNN                   &    0.037          &   0.073        &       0.042       &    0.080       &       0.108       &     0.207      &     0.100          &     0.181        \\
SASRec                  &     0.032         &     0.065       &     0.102       &    0.189       &   0.213           &   0.400        &          0.249     &    \underline{0.450}      
      \\ \midrule      
\rowcolor{gray!20}
\textbf{\ourname{}}         &       \textbf{0.072}       &      \textbf{0.151}       &     \textbf{0.127}       &  \textbf{0.215}       &   \textbf{0.263}          &      \textbf{0.437}     &       \textbf{0.285}        &       \textbf{0.452}      \\
\bottomrule
\end{tabular}
\vskip -0.1in
\end{table*}
\subsubsection{\textbf{Configurations}}

Following prior studies, recommendation performance is evaluated under the leave-one-out protocol. For each validation and test instance, 999 negative items are randomly sampled from the corresponding domain-specific item pool and ranked together with the ground-truth positive item. Performance is evaluated using Hit Ratio (HR@k) and NDCG@k.
To enable LLM-based generation of open cross-domain interactions from textual descriptions, items lacking textual metadata are excluded. To ensure data quality, users and items with fewer than 10 interactions are removed for Movie--Book, while the threshold is set to 5 for Food--Kitchen. Both overlapped and non-overlapped users are retained.
To preserve temporal and sequential characteristics, user interactions are chronologically ordered and segmented into shorter subsequences using fixed temporal windows, i.e., one month for Movie--Book and one year for Food--Kitchen. Each cross-domain subsequence is required to contain at least two items from each domain, whereas each single-domain subsequence must contain at least five items.
The resulting subsequences are partitioned into training, validation, and test sets with a ratio of 80\%/10\%/10\%, while enforcing chronological ordering to ensure that validation and test subsequences occur after the corresponding training subsequences of the same user, thereby preventing information leakage. The final interacted item in each subsequence is used as the ground-truth target.
For cross-domain intent reasoning, our \textbf{Action Intent Pair Generation} framework uses the Qwen3-4b model\footnote{https://huggingface.co/Qwen/Qwen3-4B}~\cite{bai2023qwen-1e4} to infer cross-domain intents, which are then mapped to e-commerce tags using the \textbf{BAAI/bge-m3}\footnote{https://huggingface.co/BAAI/bge-m3} text embedding model~\cite{bge-m3}, facilitating the transfer of intents across domains.

\subsection{Results on Public Datasets}

Table~\ref{tab:Main} presents the experimental results on the Movie--Book and Food--Kitchen cross-domain sequential recommendation scenarios, which were used to comprehensively assess the effectiveness of the proposed \ourname{} framework.
The experimental results indicate that the proposed \ourname{} consistently achieves superior recommendation performance across all evaluated Amazon dataset categories. In both the Movie--Book and Food--Kitchen cross-domain recommendation scenarios, \ourname{} substantially outperforms all baseline methods with respect to the two key metrics, NDCG@10 and HR@10.
Specifically, in the Movie-Book scenario, our method achieves HR@10 scores of 0.151 (Movie) and 0.215 (Book), representing improvements of 41.31\% and 0.65\% over the second-best method LLMCDSR, respectively. The performance gains are even more pronounced in the Food-Kitchen scenario, where HR@10 reaches 0.437 (Food) and 0.452 (Kitchen), surpassing LLMCDSR by 0.78\% and 3.07\%, respectively. 
These results convincingly validate the effectiveness of our approach for cross-domain recommendation tasks, particularly in realistic scenarios with substantial proportions of non-overlapping users.

Comparative analysis reveals two key findings. Conventional CDSR methods, including TPUF, $\pi$-Net, C2DSR, and MGCL, generally show limited effectiveness and may even underperform single-domain recommendation models, highlighting their reliance on overlapping users and insufficient utilization of single-domain signals. Although LLMCDSR achieves competitive performance in Food--Kitchen, its advantage is less evident in Movie--Book, whereas \ourname{} consistently maintains superior performance across both scenarios, demonstrating stronger robustness and generalization.

\subsection{Ablation Study}
\begin{table}[b]
\centering
\vskip -0.1in
\caption{Ablation Study on the Food-Kitchen Scenario.}
\vskip -0.1in
\label{tab:ablation_study}
\resizebox{\columnwidth}{!}{%
\begin{tabular}{l|cc|cc}
\toprule
\multirow{2}{*}{\textbf{Method}} & \multicolumn{2}{c|}{Food} & \multicolumn{2}{c}{Kitchen} \\
 & NDCG@10 & HR@10 & NDCG@10 & HR@10 \\
 \midrule
 \rowcolor{gray!20}
 \textbf{\ourname{}} & \textbf{0.2630} & \textbf{0.4367} & \textbf{0.2850} & \textbf{0.4521} \\
\midrule
w/o MHA & 0.2626 & 0.4200 & 0.2457 & 0.3705 \\
w/o Intent Retrieval & 0.1637 & 0.3200 & 0.1950 & 0.3564 \\
w/o Intent Tree &  0.1735 & 0.3633 & 0.1892 & 0.3469 \\
w/o  Intent Tree + Intent Retrieval   & 0.1590 & 0.2900 & 0.1031 & 0.1821 \\
\bottomrule
\end{tabular}
}

\end{table}

Table~\ref{tab:ablation_study} presents the ablation results, which demonstrate the effectiveness of each component in the \ourname{} framework.
While removing the multi-head attention mechanism (w/o MHA) causes only a slight performance decline, the Intent Retrieval and Intent Tree prove to be critical: their individual removal leads to HR@10 drops of 21-27\% and 17-23\% in Food and Kitchen domains, respectively. 
Most strikingly, simultaneously removing both components (w/o Intent Tree + Intent Retrieval) causes catastrophic performance collapse, with NDCG@10 and HR@10 plummeting by 40-64\% and 34-60\% across domains. 
This pronounced degradation further suggests that Intent Retrieval and Intent Tree function not as isolated additive modules, but as strongly coupled and indispensable components of \ourname{}. 
Intent Retrieval supports cross-domain knowledge transfer for alleviating data sparsity, while Intent Tree organizes and integrates such knowledge via hierarchical intent modeling. 
The synergy between the two is central to the effectiveness of \ourname{}, allowing it to consistently outperform existing methods in realistic non-overlapping user scenarios.

\subsection{Latency Analysis}

The analysis shows that direct invocation of Qwen3-4B leads to an inference latency of approximately \textbf{8 seconds}. In contrast, by atomizing and discretizing user behavioral intent pairs, \ourname{} preserves an effect comparable to real-time large-model reasoning while reducing latency to \textbf{20.134 ms}, achieving an approximately \textbf{400$\times$} throughput gain. This substantial reduction in inference cost is critical for real-time serving and industrial-scale recommendation deployment.

\begin{table}[t]
\centering
\caption{Category HHI Sparsity Comparison across Levels.}
\label{tab:category_sparsity_comparison}
\begin{tabular}{lcccc}
\toprule
\textbf{Model} & \textbf{Level 1} & \textbf{Level 2} & \textbf{Level 3} & \textbf{All} \\
\midrule
Base & 0.3762 & 0.0758 & 0.4434 & 0.1223 \\
\ourname{} & 0.3769 & 0.0749 & 0.4168 & 0.1201 \\
\midrule
Decrease Rate & +0.19\% & $-1.20\%$ & $-5.99\%$ & $-1.80\%$ \\
\bottomrule
\end{tabular}

\hspace{5mm}{\raggedright\footnotesize\textit{*Lower is better, a lower HHI indicates less concentration.}\par}
\end{table}
\subsection{Futher Analysis}

\subsubsection{\textbf{Product Category Analysis}}
An in-depth analysis of user data from the Kuaishou platform was conducted to investigate category sparsity across different hierarchical levels, as reported in Table~\ref{tab:category_sparsity_comparison}.
The Herfindahl–Hirschman Index (HHI) is employed to measure category concentration, with lower values reflecting a more balanced distribution of user interactions and consequently greater category diversity.
As evidenced by the results, \ourname{} consistently reduces HHI across multiple hierarchical levels, with particularly pronounced improvements at Level 2 and Level 3, where the reductions reach 1.20\% and 5.99\%, respectively.
These substantial reductions indicate that the proposed approach effectively alleviates category over-concentration and promotes broader category coverage, thereby enhancing category diversity, particularly at fine-grained and intermediate hierarchical levels.

\begin{table}[b]
\vskip -0.1in
\centering
\caption{GMV Uplift by User Activity Level.}
\vskip -0.1in
\label{tab:User}
\begin{tabular}{lcccc}
\toprule
Activity Level    & Low  & Mid & High & Ultra \\
\midrule
Lift rate & +4.32\%  & +2.46\%  & +8.04\% & +7.21\% \\
\bottomrule
\end{tabular}
\vskip -0.1in
\end{table}

\subsubsection{\textbf{User Activity Analysis}}
User activity level is closely associated with both the quantity and quality of available historical behavioral signals. To further examine model effectiveness under different levels of behavioral richness, users are stratified into four groups according to activity level, and performance is evaluated separately for each group, as reported in Table~\ref{tab:User}.
The results in Table~\ref{tab:User} show that \ourname{} consistently delivers performance gains across all user activity levels, with relative improvements ranging from +4.32\% to +8.04\%, thereby demonstrating its robustness and broad effectiveness.
Notably, for low-activity users, the observed +4.32\% improvement indicates that the LLM-generated user behavior intent tree effectively alleviates the cold-start issue by transforming sparse cross-domain interaction signals into more comprehensive behavioral representations, thereby compensating for limited historical data. For high-activity and ultra-activity users (+8.04\% and +7.21\%, respectively), the pronounced gains further verify the effectiveness of the target-item-conditioned reasoning compression mechanism in modeling long behavioral sequences.

By selectively suppressing irrelevant noise and retaining behaviorally salient signals from ultra-long interaction histories, the proposed approach enables the model to capture fine-grained and context-dependent user preferences without being hindered by the excessive volume of historical behaviors.

\subsection{Online A/B Test}
\begin{table}[t]
\centering
\caption{Industry Online A/B Test Results.}
\label{tab:Industry}
\begin{tabular}{lcccc}
\toprule
Metric    & Paid Order Count  & GMV & GPM & OPM\\
\midrule
Lift rate & +1.043\%  & +3.446\%  & +3.662\% & +1.254\% \\
\bottomrule
\end{tabular}
\end{table}

The proposed \ourname{} framework has been deployed in the e-commerce short-video scenario on the Kuaishou platform.
Rigorous online A/B tests were conducted during a online deployment to validate the effectiveness of the proposed model. As reported in Table~\ref{tab:Industry}, \ourname{} achieves significant improvements over the previous online baseline in this setting.
Substantial and consistent improvements are observed across several core business metrics. Specifically, Paid Order Count increases by +1.043\%, indicating enhanced conversion performance; GMV and GPM improve by +3.446\% and +3.662\%, respectively, demonstrating stronger monetization capacity; and OPM rises by +1.254\%, suggesting improved overall profit efficiency. Collectively, these results provide strong evidence for the practical utility of \ourname{} in real-world industrial recommendation scenarios.

\section{Conclusion}
\ourname{} is a novel LLM-driven framework for cross-domain intent modeling and reasoning-oriented retrieval, specifically designed to address the challenges posed by noisy, massive, and heterogeneous multi-domain user behavior sequences in industrial-scale e-commerce recommendation systems. 
By jointly integrating user attributes and cross-domain interaction logs, \ourname{} constructs structured Action-Intent Pairs and dynamically builds a preference-aware, weighted Intent Tree, enabling fine-grained semantic reasoning and target-aware retrieval to extract high-evidence User Intent Chains for more accurate CTR prediction and ranking.
Notably, we adopt an offline-online hybrid strategy that pre-generates atomic intents offline using large language models and assembles them dynamically during online serving. This design achieves functionality equivalent to real-time LLM reasoning while eliminating the prohibitive latency of online LLM inference, resulting in millisecond-level serving latency and an approximately \textbf{400$\times$} inference throughput gain compared to direct model invocation. 
Through effective noise filtering, behavior sequence compression, and efficient intent aggregation, \ourname{} significantly improves both modeling efficiency and recommendation quality. Extensive experiments on public benchmarks, together with large-scale industrial A/B tests, demonstrate the effectiveness and real-world applicability of our framework, yielding consistent online gains in key business metrics, including \textbf{GMV (+3.446\%)} and \textbf{GPM (+3.662\%)}.

\begin{acks}
This work is supported by Kuaishou Technology. 
The research described in this paper has been partially supported by the General Research Funds from the Hong Kong Research Grants Council (project No. PolyU 15200023, 15206024, and 15224524), Hong Kong Research Grants Council’s Theme-based Research Scheme (No. T43-513/23-N), Hong Kong Research Grants Council’s Research Impact Fund (No. R1015-23), Hong Kong Research Grants Council’s Collaborative Research Fund (No. C1043-24GF), and Internal research funds from Hong Kong Polytechnic University (project no. P0059586, P0042693, P0048625, and P0051361). This work was supported by computational resources provided by The Centre for Large AI Models (CLAIM) of The Hong Kong Polytechnic University.

\end{acks}


\bibliographystyle{ACM-Reference-Format}
\balance
\bibliography{citation}

\end{document}